\documentclass[journal]{IEEEtran}
\usepackage{subfigure}
\usepackage{graphicx}
\usepackage{epsfig,latexsym,amsmath,graphics}
\usepackage{amssymb,dsfont,amsthm}
\usepackage{cite,color}
\usepackage{url}
\usepackage{booktabs}
\usepackage{algorithm,algorithmicx}
\usepackage{algpseudocode}
\usepackage{stfloats}
\usepackage{verbatim}
\usepackage{bm}
\usepackage{stfloats}
\usepackage{multirow}
\usepackage{ulem}
\usepackage{multirow}

\begin{document}
	%
	\title{Detection Capability Comparison Between Intensity Detection and Splitting Detection for Rydberg-Atomic Sensors}
	%
	%
	%
		%
	
	\author{Hao Wu, Xinyuan Yao, Rui Ni, Chen Gong and Kaibin Huang
		\thanks{This work was supported by National Natural Science Foundation of China	under Grant 62331024 and 62171428.}
		\thanks{Hao Wu, Xinyuan Yao, Chen Gong are with the School of Information Science and Technology in University of Science and Technology of China, Email address: \{wuhao0719, yxy200127\}@mail.ustc.edu.cn, cgong821@ustc.edu.cn.
			
		Rui Ni is with Huawei Technology, Email address: raney.nirui@huawei.com.
		
		K. Huang is with the Department of Electronic and Electrical Engineering in the University of Hong Kong, Hong Kong, Email address: huangkb@eee.hku.hk.

}
	}

\maketitle

\begin{abstract}
Rydberg atomic quantum receivers have been seen as novel radio frequency measurements and the high sensitivity to a large range of frequencies makes it attractive for communications reception. However, their unique physical characteristics enable two fundamental signal readout schemes: intensity-based detection and splitting-based detection. The former measures the electric fields through laser intensity, while the latter utilizes Autler-Townes splitting. In this work, we systematically categorize and model existing signal readout methods, classifying them into these two paradigms. Then, we derive the maximum likelihood estimation procedures and corresponding Cramér-Rao lower bounds (CRLB) for each detection modality. Through the analysis of the CRLB, we propose strategy for both readout schemes to enhance sensitivity and minimize estimation variance: acquiring data in regions with maximal slope magnitudes. While this approach has been implemented in intensity-based detection (e.g., superheterodyne schemes), its application to splitting-based detection remains unexplored. Implementation of non-uniform frequency scanning, with preferential sampling at regions exhibiting maximum peak slopes combined with our proposed maximum likelihood splitting estimation method, achieves significantly reduced estimation variance compared to conventional polynomial fitting. The comparative analysis reveals the optimal detection performance of the two detection schemes. This work also contributes to enhancing the accuracy of microwave calibration. Numerical results reveal that both fundamental signal readout methods achieve lower estimation variance based on our proposed maximum likelihood estimation approach.

\end{abstract}

\section{Introduction}
Recently, Rydberg atoms have emerged as a novel platform for electric field sensing, enabling direct International System of Units (SI)-traceable and self-calibrated measurements\cite{anderson2021self,song2017quantum}. The measured splitting only depends on the field strength, Planck’s constant, and dipole moment of the Rydberg transition, which can be reliably calculated. Rydberg atoms in highly excited states with one or more electrons of large principal quantum numbers are sensitive to electric fields, very suitable to manufacture atom-based sensor for detecting communication signals. 

In work \cite{gong2025rydberg}, the authors summarize eight fundamental receiver architectures, including the standard and superheterodyne schemes, and propose single-input single-output and multiple-input multiple-output models based on Rydberg atoms. Furthermore, all the schemes and models realize RF-to-optical conversions based on two types of atom-light interactions, electromagnetically induced transparency (EIT) and Autler–Townes splitting (ATS).  Informed by the principles of EIT and ATS, the receiving and signal readout methods can be categorized into two distinct detection methodologies: intensity detection (ID) and splitting detection (SD). The former encompasses all readout techniques that rely on the correlation between transmitted laser intensity and RF power, while the latter is concerned with readout methods that analyze the relationship between the distance of double peaks and RF power.

The intensity-based detection approach eliminates the need for laser frequency scanning, as the detected optical signal intensity is directly used for subsequent signal processing and analysis. All detection schemes, including standard, superheterodyne, and frequency scanning-free configurations, are fundamentally categorized as intensity detection modalities, distinguished by their respective signal processing methodologies for detected optical waveforms. As a simplified signal readout methodology, intensity-based detection has been extensively employed in polarization measurement \cite{sedlacek2013atom}, angle-of-arrival estimation \cite{robinson2021determining,mao2023digital}, subwavelength imaging \cite{holloway2017atom,downes2020full}, near-field antenna pattern characterization \cite{simons2019embedding,shi2023new}, and multi-frequency signal recognition \cite{liu2022deep,zhang2022rydberg,zhang2024image}. 

The detection of frequency splitting is fundamentally grounded in the ATS phenomenon, wherein RF field strength quantification is achieved through precise measurement of the spectral bifurcation between dual resonance peaks in the ATS profile. Different from nonlinear intensity detection, under specific experimental conditions, the observed ATS phenomenon demonstrates a distinct linear correlation with the amplitude of the externally applied RF field, establishing a quantifiable relationship critical for field strength determination\cite{holloway2017electric}. The frequency splitting-based detection has long served as a classical and conventional signal detection methodology, widely employed in precision measurements requiring self-calibration, such as microwave calibration\cite{holloway2014sub,simons2018electromagnetically,cui2025towards}.

Currently, the reception scheme based on intensity detection combined with RF heterodyne demonstrates superior sensitivity performance compared to splitting detection methods\cite{schlossberger2024rydberg}. However, the underlying mechanisms responsible for this phenomenon remain fragmented and lack a unified interpretation. For the intensity detection based on RF heterodyne, previous works \cite{wu2022rydberg,jing2020atomic} demonstrate that the gain is governed by the slope of the system response function. Slope-based optimization can consequently enhance the detection sensitivity. For ATS-based splitting detection schemes, the EIT linewidth (2-5 MHz under low-power conditions) fundamentally limits the spectral resolution, preventing accurate field detection for splittings narrower than the linewidth. This constraint governs both the sensor’s minimum detectable field and sensitivity, with work \cite{holloway2017electric} establishing a practical threshold: the Rabi frequency of RF field $\Omega_{RF}>2 \times$EIT linewidth for SI-traceable accuracy\cite{holloway2022rydberg}.

From the perspective of signal detection, intensity and splitting detection constitute two distinct estimation approaches for the same target variable, i.e., electric field intensity. Mathematically, the sensitivity disparity between these structures in work \cite{schlossberger2024rydberg} originates from differences in their respective estimation variances. Given the universal nature of this limitation in Rydberg-based perception, the analytical conclusions are equally applicable to alternative configurations, including optical homodyne structures\cite{schlossberger2024rydberg}.

In this work, we establish a systematic classification framework for signal readout methodologies, categorizing them into two fundamental paradigms: intensity detection and splitting detection. Specifically, we exemplify two canonical intensity detection models: intensity direct detection and intensity superheterodyne detection. For splitting detection, we differentiate between two model classes based on prior knowledge of the peak function. The peak function is fully known a priori (univariate estimation) and the peak function is partially known a priori (multivariate estimation). Leveraging these modeled readout schemes and Gaussian noise, we derive the maximum likelihood estimation (MLE) procedures and the corresponding Cramér-Rao lower bounds (CRLB) for each detection modality. Through the CRLB analysis, we propose strategy for both readout schemes to enhance sensitivity and minimize the estimation variance: acquiring data in regions with maximal slope magnitudes. While this approach has been implemented in intensity detection (e.g., intensity superheterodyne schemes), its application to splitting-based detection remains unexplored. Implementation of non-uniform frequency scanning, with preferential sampling at regions exhibiting maximum peak slopes combined with our proposed maximum likelihood splitting estimation method, achieves significantly reduced estimation variance compared to conventional polynomial fitting. This work also contributes to enhancing the accuracy of microwave calibration. Numerical results reveal that both fundamental signal readout methods achieve lower estimation variance based on our proposed maximum likelihood estimation approach.

The remainder of this paper is organized as follows. In Section \ref{T2}, we introduce the basic models required, including four-level Rydberg atoms physical model and Gaussian noise model. In Section \ref{T3} and Section \ref{T4}, we systematically model and analyze the Cramér-Rao lower bounds for both intensity detection and splitting detection schemes, respectively, and derive their corresponding maximum likelihood estimation methods. In Section \ref{T5}, we conduct a systematic comparison between these detection schemes based on CRLB. In Section \ref{T6}, we show the numerical results. Finally, Section \ref{T7} concludes this work.

%
%
%
%


\section{Preliminary Quantum System Model}\label{T2}

\subsection{Theoretical Physics Model}\label{T22}

\begin{figure*}
\centering
\includegraphics[width=0.89\textwidth]{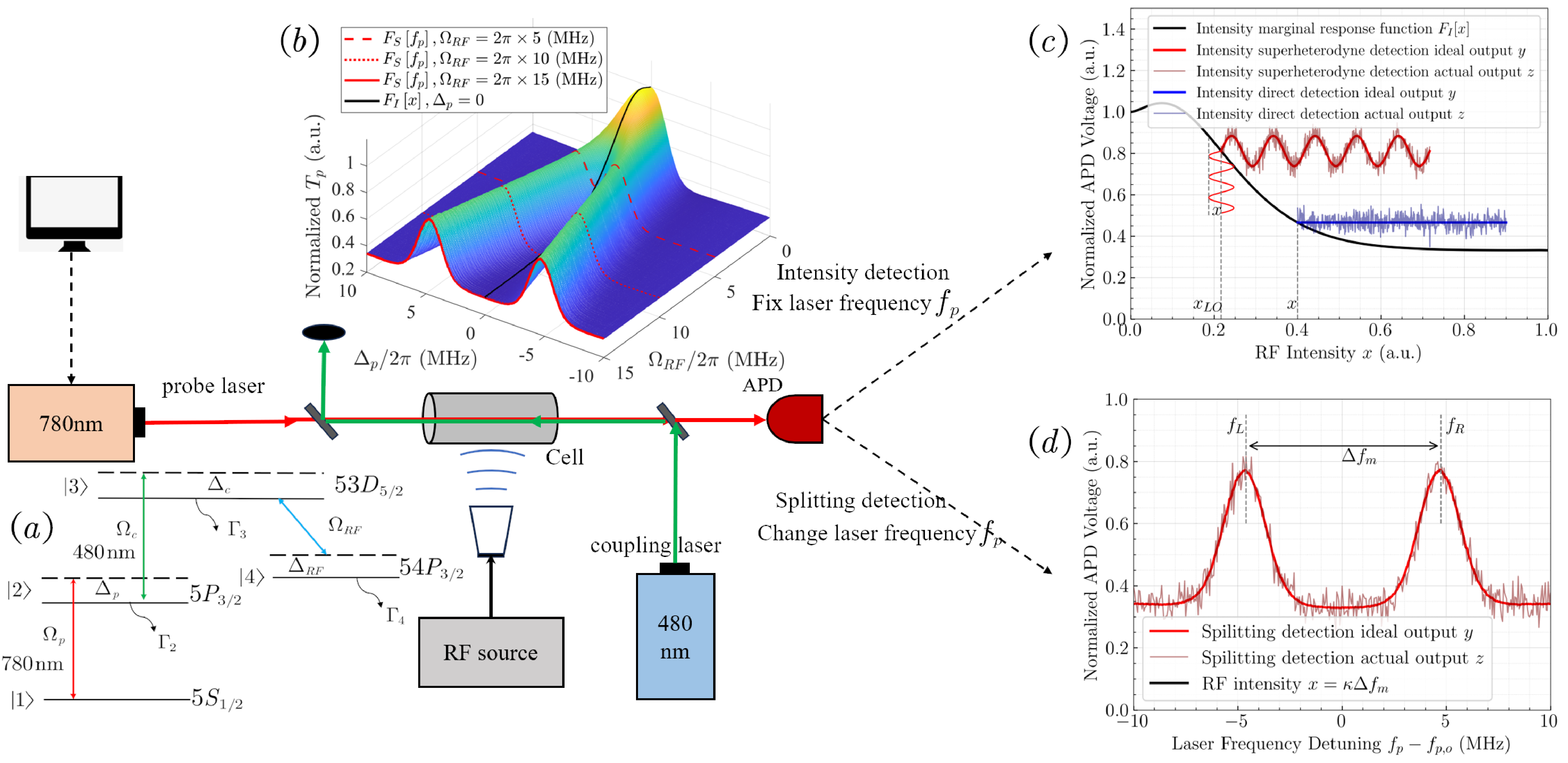}
\caption{(a) The diagram for the four-level scheme. (b) Normalized $\mathcal{G}\left[ x,f_p \right]=\mathcal{G}\left[ \frac{\hbar\Omega _{RF}}{\mu _{RF}},f_{p,o}+\Delta_p \right]$ with respective to detuning $\Delta_{p}$ and RF intensity $x=\Omega_{RF}\frac{\hbar}{\mu_{RF}}$ under $\Omega_p =2\pi \times 2$ MHz and $\Omega_c =2\pi \times 4$ MHz. (c) Intensity detection output. (d) Splitting detection output}
\label{diagram}
\end{figure*}

The typical Rydberg four-level ladder system is shown in Fig. \ref{diagram}(a). The probe and coupling laser beams excite the atoms from state $\left| 1 \right> $ to state $\left| 2 \right> $ and from state $\left| 2 \right> $ to state $\left| 3 \right>$, respectively. The RF field couples Rydberg state $\left| 3 \right>$ and state $\left| 4 \right> $. Let $\Delta _p$, $\Delta _c$, and $\Delta _{RF}$ denote the detunings for the probe laser, couple
laser and RF electric field, respectively. Let $\Omega _p$, $\Omega _c$, and $\Omega _{RF}$ denote Rabi frequencies associated with the probe laser, couple laser and RF electric field, respectively. We have
\begin{equation}\label{OE}
\begin{aligned}
\Omega _{p}&=|E_{p}|\frac{\mu _{p}}{\hbar },\\
\Omega _{c}&=|E_{c}|\frac{\mu _{c}}{\hbar },\\
\Omega _{RF}&=|E_{RF}|\frac{\mu _{RF}}{\hbar },
\end{aligned}
\end{equation}	
where $|E_{p}|$, $|E_{c}|$ and $|E_{RF}|$ are the magnitudes of the electric-field of the probe laser, coupling laser, and RF source, respectively. Let $\mu _{p}$, $\mu _{c}$ and $\mu _{RF}$ denote the atomic dipole moments corresponding to $\left| 1 \right> -\left| 2 \right> $, $\left| 2 \right> -\left| 3 \right> $ and $\left| 3 \right> -\left| 4 \right>$, respectively. Let $\hbar$ denote the reduced planck constant. Assume that the decay rates are $\Gamma_{2}$ for state $\left| 2 \right>$, $\Gamma_{3}$ for state $\left| 3 \right>$, and $\Gamma_{4}$ for state $\left| 4 \right>$\cite{holloway2017electric}.

The Lindblad master equation of above atomic system is given as follows\cite{auzinsh2010optically}:
\begin{equation}\label{Lindblad master equation}
\begin{aligned}
\mathbf{\bm{\dot\rho}}=\frac{\partial \bm{\rho }}{\partial t}=-\frac{i}{{\hbar }}\left[ \bm{H},\bm{\rho } \right] +\bm{\mathcal{L}},
\end{aligned}
\end{equation}
where $\bm\rho$ denotes the density matrix of the atomic ensemble and $\bm{\mathcal{L}}$ denotes Lindblad operator.
The interaction between RF signals and Rb atoms affects component $\rho_{21}$ of the $4\times4$ matrix $\mathbf{\bm{\rho}}$. Element $\rho_{21}$ in the second row and first column of matrix $\bm{\rho}$ is related to the transmittance of atom vapor cell, which finally affects the output probe laser power. The corresponding Hamiltonian in the rotating wave approximation frame is given by

\begin{small}
\begin{equation}
\begin{aligned}
	\boldsymbol{H}=\frac{{\hbar }}{2}\left( \begin{matrix}
		0&		\Omega _p&		0&		0\\
		\Omega _p&		-2\Delta _p&		\Omega _c&		0\\
		0&		\Omega _c&		-2\left( \Delta _p+\Delta _c \right)&		\Omega _{RF}\\
		0&		0&		\Omega _{RF}&		-2\left( \Delta _p+\Delta _c+\Delta _{RF} \right)\\
	\end{matrix} \right) .
\end{aligned}
\end{equation}
\end{small}
The relationship between transmittance $T_p$ of the atomic
medium and $\rho_{21}$ is given by
\begin{equation}\label{Tp}
\begin{aligned}
T_p=e^{\frac{2N_0\mu _{p}^{2}kL}{\epsilon {\hbar }\Omega _p}\Im \left( \rho _{21} \right)},
\end{aligned}
\end{equation}
where $N_0$ is the total density of atoms and $\epsilon$ is the permittivity in vacuum. Let $L$ denote the length of atomic cell and $k=2\pi/\lambda_p$ denote the wave vector of the probe laser. Let $\Im (\rho_{21})$ characterize the imaginary part of $\rho_{21}$.

In addition, in the case of thermal atom model, Doppler effect caused by the atomic thermal motion. The Doppler averaged density matrix element $\rho_{21D}$ can be expressed as\cite{holloway2017electric}
\begin{equation}
\begin{aligned}
\rho _{21D}=\frac{1}{\sqrt{\pi}u}\int_{3u}^{3u}{\rho _{21}\left( \Delta _{p}^{'},\Delta _{c}^{'} \right) e^{-\frac{v^2}{u^2}}dv},
\end{aligned}
\end{equation}
where $u =\sqrt{k_BT/m}$, $m$ is the atom's mass, $k_B$ is the Boltzmann constant, and $T$ is the thermodynamic temperature. At room temperature, $T = 303.15 \text{K}$, due to atomic thermal motion, the probe and coupling light detuning $\Delta _{p}^{'}$ and $\Delta _{c}^{'}$ should be modified by the following

\begin{equation}
\begin{aligned}
\Delta _{p}^{'}=\Delta _p-\frac{2\pi}{\lambda _p}v,
\\
\Delta _{c}^{'}=\Delta _c+\frac{2\pi}{\lambda _c}v.
\end{aligned}
\end{equation}

Due to the gas transmittance $T_p$ is affected by laser detuning $\Delta_{p}$ and microwave intensity $\left| E_{RF} \right|$, resulting in changes in the optical power transmitted through the atomic cell. We define $\mathcal{G}\left[ x,f_p \right]$ as the APD joint response function, representing the output voltage for given microwave intensity $x=\left| E_{RF} \right|$ and probe laser detuning $\Delta_p=f_p-f_{p,o}$, where $f_p$ and $f_{p,o}$ are the probe laser frequency and resonant frequency of laser-coupled atomic transition, respectively.

Figure \ref{diagram}(b) presents a three-dimensional visualization of parameter $\mathcal{G}\left[ x,f_p \right]=\mathcal{G}\left[ \frac{\hbar\Omega _{RF}}{\mu _{RF}},f_{p,o}+\Delta_p \right]$, generated using the established model and parameter configuration from reference \cite{holloway2017electric}. Although the cavity transmittance $T_p$ in Eq. (\ref{Tp}) depends on multiple physical parameters (e.g., $\Omega_{p}$), our analysis focuses on fundamental characteristics rather than system response optimization. We therefore adopt $\frac{2N_0\mu _{p}^{2}kL}{\epsilon {\hbar }\Omega _p}=1,$ and normalize the transmittance and $\mathcal{G}$ throughout this study. All subsequent analyses are performed using this normalized results for consistency.


\subsection{Noise Model}\label{T23}

Define $y$ as the ideal avalanche photon diode (APD) output voltage in the absence of noise, representing the fundamental signal response, i.e., $y=\mathcal{G}\left[ x,f_p \right]$ for given $x$ and $f_p$. In actual atomic systems, the measured APD voltage $z$ suffer from different sources of noise $n$. Nevertheless, whether originating from quantum system or receiver APD, these noise components predominantly exhibit Gaussian characteristics \cite{wu2025off}. In this work, we adopt the approximation that all noise contributions are signal-independent additive white Gaussian noise $n_0$. Thus, the actual APD voltage $z$ can be expressed as
\begin{equation}
\begin{aligned}
z=y+n_0,
\end{aligned}
\end{equation}
where the variance of noise $n_0$ is $\sigma _{0}^{2}$.


\section{Intensity Detection Model}\label{T3}

Since that the laser parameters are fixed (including detuning $\Delta_p$) throughout signal acquisition in the detection process. This approach encompasses all readout methodologies that rely on the correlation between transmitted laser power and true RF strength $x$. Consequently, we characterize the detection model through its intensity marginal response function $F_I\left[ x \right]$, corresponding to a cross-section of the APD joint response function $\mathcal{G}\left[ x,f_p \right]$ given laser detuning $\Delta_p$, given by
\begin{equation}
\begin{aligned}
F_I\left[ x \right] =\left. \mathcal{G}\left[ x,f_p \right] \right|_{f_p=f_{p,o}+\Delta _p},
\end{aligned}
\end{equation}
where the true electric field strength $x=\left| E_{RF} \right|$ is encoded in the transmitted laser intensity, and quantitatively measured through the ideal APD output voltage. As illustrated in Fig. \ref{diagram}(b), the black solid curve represents the $F_I\left[ x \right]$ function at zero detuning $\Delta_{p}=0$. 

Currently, there are two predominant detection schemes: intensity direct detection (IDD) and intensity superheterodyne detection (ISD). While exhibiting distinct operational architectures, both methods fundamentally rely on intensity marginal response function $F_I\left[ x \right]$ under fixed laser parameters, differing primarily in their post-detection signal processing methodologies. Figure \ref{diagram}(c) illustrates the outputs of the two architectures for signal $x$.

\subsection{Intensity Direct Detection}\label{T31}
\subsubsection{Detection Model}\label{T311}

Since there is no need for a frequency scanning process, the receivers adopting direct detection can directly calculate true electric field strength $x$ based on the ideal APD voltage $y=F_I\left[ x \right]$ and inverse function $F^{-1}_I[y]$. However, due to the presence of noise, the collected noisy APD data $z$ follow Gaussian distributions with mean $y$ and variance $\sigma _{0}^{2}$, resulting in the error for the estimated electric field strength $\tilde{x}=F^{-1}_I[z]$. The estimated electric field strength $\tilde{E}_{IDD}$ based on intensity direct detection model can be shown as the function of $y$ and noise $n_0$, given by

\begin{equation}
\begin{aligned}
\left| \tilde{E}_{IDD} \right|&=\tilde{x}=F_{I}^{-1}\left[ z \right] 
\\
&=F_{I}^{-1}\left[ y+n_0 \right] 
\\
&\overset{1}{=}F_{I}^{-1}\left[ y \right] +\frac{\partial F_{I}^{-1}\left[ y \right]}{\partial y}n_0+\mathcal{O}\left( n_0 \right) 
\\
&\overset{2}{=}x+\frac{1}{F_{I}^{\left( 1 \right)}\left[ x \right]}n_0+\mathcal{O}\left( n_0 \right) ,
\end{aligned}
\end{equation}
where $F_{I}^{\left( 1 \right)}\left[ x \right]$ and $F_{I}^{-1}\left[ y \right]$ are the first derivative and inverse function of the function $y=F_{I}\left[ x \right]$, respectively. In $\overset{1}{=}$, we assume that noise $n_0$ is relatively small compared to the ideal APD voltage $y$. Applying a first-order Taylor expansion to simplify $F_I^{-1}\left[ y+n \right] $ and ultimately obtain an approximate linear model. In $\overset{2}{=}$, $x=F_{I}^{-1}\left[ y \right]$ is based on the definition of inverse function and the derivative is equal to the reciprocal of the derivative of the original function. Specifically, the electric field strength $\left| \tilde{E}_{IDD} \right|$ estimated through direct detection can be expressed as the true electric field strength $x$ superimposed with noise that has been either amplified or attenuated by $\frac{1}{F_{I}^{\left( 1 \right)}\left[ x \right]}$. 

\subsubsection{Maximum Likelihood Estimation}\label{T312}

Thus, for low power noise, the direct detection can be regarded as a linear Gaussian noise model, where the core challenge in electric field estimation $\tilde{x}$ reduces to reconstructing the ideal APD voltage $\tilde{y}$ from sampled voltage data $\boldsymbol{z}$, given by
\begin{equation}
\begin{aligned}
\boldsymbol{z}\xrightarrow{\text{MLE}}\tilde{y}\xrightarrow{F_{I}^{-1}}\tilde{x},
\end{aligned}
\end{equation}
where maximum likelihood estimation (MLE) is employed to reconstruct the ideal voltage $\tilde{y}$, subsequently inferring the electric field strength $\tilde{x}$ through inverse function mapping $\tilde{x}=F_{I}^{-1}\left[\tilde{y} \right] $. Since all samples are independent, the joint probability density function (PDF) of $N$ observations $\boldsymbol{z}=\left[ z_1,z_2,\cdots ,z_N \right] $ can be written as
\begin{equation}
\begin{aligned}
p\left( \boldsymbol{z};y \right) =\exp \left( -\frac{1}{2}\sum_{i=1}^N{\frac{\left( z_i-y \right) ^2}{\sigma_0 ^{2}}} \right) \times \prod_{i=1}^N{\frac{1}{\sqrt{2\pi \sigma_0 ^{2}}}}.
\end{aligned}
\end{equation}

Then, the maximum likelihood estimation method, derived from optimization of the likelihood PDF, can be equivalently transform into a log-likelihood maximization problem, given by
\begin{equation}
\begin{aligned}
\tilde{y} &=\mathop{\arg\max}\limits_{y} \ln p\left( \boldsymbol{z};y \right)
\\
&=\mathop{\arg\min}\limits_{y} \sum_{i=1}^N{\frac{\left( z_i-y \right) ^2}{\sigma _{0}^{2}}}.
\end{aligned}
\end{equation}

The estimated $\tilde{y}$ can be determined by evaluating $\frac{\partial}{\partial y}\ln p\left( \boldsymbol{z};y \right)=0$. We have $\tilde{y}=\frac{1}{N}\sum_{i=1}^N{z_i}$. Finally, the estimated electric field strength $\tilde{E}_{IDD}$ based on direct detection model can be expressed as
\begin{equation}
\begin{aligned}
\left| \tilde{E}_{IDD} \right|&=\tilde{x}=F_{I}^{-1}\left[ \tilde{y}\right] =F_{I}^{-1}\left[ \frac{1}{N}\sum_{i=1}^N{z_i}\right] .
\end{aligned}
\end{equation}

\subsubsection{Cramér-Rao Lower Bounds}\label{T313} \textcolor{white}{1}

\normalem{\emph{Theorem 1:}} Under $N$ sampling data and noise $\sigma_{0}$, the CRLB of intensity direct detection is given as follows,
\begin{equation}
	\begin{aligned}
		\text{CRLB}_{IDD} &= \frac{\sigma _{0}^{2}}{N}\left( \frac{1}{F_{I}^{\left( 1 \right)}\left[ x \right]} \right) ^2.
	\end{aligned}
\end{equation}

\normalem{\emph{Proof:}} Follow the definition of CRLB.

\subsection{Intensity Superheterodyne Detection}\label{T32}
\subsubsection{Detection Model}\label{T321}
Different from the intensity direct detection model, APD voltage $z\left( t\right) $ of superheterodyne detection model can expressed as
\begin{equation}
\begin{aligned}
z(t)&=y(t)+n_0(t)
\\
&=F\left[ x_{LO} \right] +F^{\left( 1 \right)}\left[ x_{LO} \right] x\cos \left( 2\pi \Delta f_st+\phi \right) +n_0(t),
\end{aligned}
\end{equation}
where APD voltage $z$ contains DC component $F\left[ x_{LO} \right]$, AC component $F^{\left( 1 \right)}\left[ x_{LO} \right] x\cos \left( 2\pi \Delta f_st+\phi \right)$, and noise $n_0(t)$. $x_{LO}$ and $x$ are local oscillator signal strength and receiving microwave signal strength, respectively. $\Delta f_s=f_{LO}-f_s$ and $\phi$ are the frequency and phase difference between the local oscillator signal and the microwave signal.

\subsubsection{Maximum Likelihood Estimation}\label{T322}
The signal readout scheme in superheterodyne structures relies on the relationship between the intermediate-frequency amplitude and signal strength, necessitating complete acquisition of at least one full cosine period. Thus, each sampling operation within a cosine period can be regarded as an independent measurement of the signal intensity $x$. We can analyze the measurements within a single period as a representative case. When sampling at intervals of $T_{SD,2}=\frac{1}{N_{SD,2}\Delta f_s}$, since each cosine period contains $N_{SD,2}$ discrete sampling points, the sampled APD voltage $z_k$ can be expressed as

\begin{equation}
\begin{aligned}
z_k&=z\left( kT_{SD,2} \right) 
\\
&=F_I\left[ x_{LO} \right] +F_I^{\left( 1 \right)}\left[ x_{LO} \right] x\cos \left( 2\pi \frac{k}{N_{SD,2}}+\phi \right) +n_0,
\end{aligned}
\end{equation}
where $\phi$ is the phase; $k=1, 2, \cdots, N_{SD,2}$ indexes the $N_{SD,2}$ discrete sampling points. Adopting discrete fourier transform and following by absolute value operation, we can obtain the intensity of the frequency component $\Delta f_s$ within one period, given by

\begin{equation}
\begin{aligned}
&\frac{N_{SD,2}}{2}F_I^{\left( 1 \right)}\left[ x_{LO} \right] \tilde{x}_i=\left| \sum_{k=1}^{N_{SD,2}}{z_ke^{-j\frac{2\pi}{N_{SD,2}}\left( k-1\right) }}\right| 
\\
&=\left| F_I \left[ x_{LO} \right] \sum_{k=1}^{N_{SD,2}}{e^{-j\frac{2\pi\left( k-1\right) }{N_{SD,2}} }}+\sum_{k=1}^{N_{SD,2}}{n_0e^{-j\frac{2\pi\left( k-1\right) }{N_{SD,2}} }}\right. 
\\
&\left. +F_I^{\left( 1 \right)}\left[ x_{LO} \right] x\sum_{k=1}^{N_{SD,2}}{\cos \left( 2\pi \frac{k}{N_{SD,2}}+\phi \right) e^{-j2\pi \frac{k-1}{N_{SD,2}}}}\right| 
\\
&=\left| N_{SD,2}\frac{F_I^{\left( 1 \right)}\left[ x_{LO} \right] x}{2}e^{j\phi}+\frac{N_{SD,2}}{2}\tilde{n}_0\right| ,
\end{aligned}
\end{equation}
where $\tilde{n}_0=\frac{2}{N_{SD,2}}\sum_{k=0}^{N_{SD,2}-1}{n_0e^{-j\frac{2\pi}{N_{SD,2}}k}}$ is complex Gaussian noise with variance $\mathbb{D}\left[\tilde{n}_0 \right]=4\mathbb{D}\left[n_0 \right] =4\sigma_0^2$. Then, the strength estimate $\tilde{x}_i$ based on data $z_k$ from the $i$-th cosine period can be expressed as

\begin{equation}
\begin{aligned}
\tilde{x}_i&=\frac{2}{F_I^{\left( 1 \right)}\left[ x_{LO} \right] N_{SD,2}}
\\
&\ \ \ \ \times \left| N_{SD,2}\frac{F_I^{\left( 1 \right)}\left[ x_{LO} \right] x}{2}e^{j\phi}+\frac{N_{SD,2}}{2}\tilde{n}_0 \right|
\\
&\overset{1}{=}\left| x+\frac{1}{F_I^{\left( 1 \right)}\left[ x_{LO} \right]}e^{-j\phi}\tilde{n}_0 \right|
\\
&\overset{2}{=}x+\Re \left[ \frac{1}{F_I^{\left( 1 \right)}\left[ x_{LO} \right]}\tilde{n}_0 \right] 
\\
&\overset{3}{=}x+\frac{\sqrt{2}}{F_I^{\left( 1 \right)}\left[ x_{LO} \right]}n_0,
\end{aligned}
\end{equation}
where $x$, $F^{\left( 1 \right)}\left[ x_{LO} \right]$ and $N_{SD,2}$ are real number and $\tilde{n}$ is complex Gaussian noise in $\overset{1}{=}$. In $\overset{2}{=}$, $e^{-j\phi}\tilde{n}_0$ and $\tilde{n}_0$ follow identical complex Gaussian distributions. Similar to Eq. (9), we consider that the noise $\tilde{n}_0$ is relatively small compared to the ideal amplitude $F_I^{\left( 1 \right)}\left[ x_{LO} \right]x$, thereby allowing neglect of the imaginary component's contribution to the absolute value from complex noise $\tilde{n}_0$. In $\overset{3}{=}$, the real component of complex Gaussian noise $\tilde{n}_0$ remains Gaussian-distributed, but with half the original power, which allows the approximation $\Re \left[\tilde{n}_0 \right] =\sqrt{2}n_0$.

Therefore, the strength estimation $\tilde{E}_{ISD}$ derived from $N_{SD,2}$ sampling points based on the $i$-th cosine period can be modeled as a linear superposition of the true signal component $x$ and equivalent measurement noise $n_{ISD}$, given by
\begin{equation}
\begin{aligned}
\left| \tilde{E}_{ISD} \right|&=\tilde{x}_i=x+n_{ISD}
\\
&=x+\frac{\sqrt{2}}{\sqrt{N_{SD,2}}F_{I}^{\left( 1 \right)}\left[ x_{LO} \right]}n_0,
\end{aligned}
\end{equation}
where the equivalent measurement noise is the function of the local oscillator signal $x_{LO}$ rather than $x$, with a constant variance $\sigma _{ISD}^{2}=\frac{2\sigma _{0}^{2}}{N_{SD,2}\left( F_I^{\left( 1 \right)}\left[ x_{LO} \right] \right) ^2}$.

Generally, we consider sampling $ N_{SD,2}$ points per cosine period over $N_{SD,1}$ complete periods, yielding a total of $N=N_{SD,1}\times N_{SD,2}$ sample points $\boldsymbol{z}$, which is given by

\begin{equation}
\begin{aligned}
\boldsymbol{z}=\underset{1}{\underbrace{\left[ z_1,z_2,\cdots ,z_{N_{SD,2}} \right. }},\underset{2}{\underbrace{z_{N_{SD,2}+1},\cdots ,z_{2N_{SD,2}}}},\cdots 
\\
,\underset{N_{SD,1}}{\underbrace{\left. z_{\left( N_{SD,1}-1 \right) \times N_{SD,2}+1},\cdots ,z_{N_{SD,1}\times N_{SD,2}} \right] }}.
\end{aligned}
\end{equation}

Since each cosine period is statistically independent, the corresponding estimates $\tilde{x}_i$ derived from individual periods are likewise mutually independent. The joint PDF of $N_{SD,1}$ observations under a total of $N=N_{SD,1}\times N_{SD,2}$ sample points can be written as
\begin{equation}
\begin{aligned}
p\left( \boldsymbol{\tilde{x}};x \right) =\exp \left( -\frac{1}{2}\sum_{i=1}^{N_{SD,1}}{\frac{\left( \tilde{x}_i-x \right) ^2}{\sigma _{ISD}^{2}}} \right) \times \prod_{i=1}^{N_{SD,1}}{\frac{1}{\sqrt{2\pi \sigma _{ISD}^{2}}}},
\end{aligned}
\end{equation}
where $\boldsymbol{\tilde{x}}=\left[ \tilde{x}_1,\tilde{x}_2,\cdots ,\tilde{x}_{N_{SD,1}} \right] $. 

The maximum likelihood estimation for intensity superheterodyne detection is established based on $\left| \tilde{E}_{ISD} \right|=\mathop{\arg\max}\limits_{x} \ln p\left( \boldsymbol{\tilde{x}};x \right)$. By evaluating $\frac{\partial}{\partial x}\ln p\left( \boldsymbol{\tilde{x}};x \right)=0$, we have
\begin{equation}
\begin{aligned}
\frac{\partial}{\partial x}\ln p\left( \boldsymbol{\tilde{x}};x \right) &=\frac{1}{2}\sum_{i=1}^{N_{SD,1}}{\frac{2\left( \tilde{x}_i-x \right)}{\sigma _{ISD}^{2}}}=0,
\end{aligned}
\end{equation}
where $\left| \tilde{E}_{ISD} \right|=\frac{1}{N_{SD,1}}\sum_{i=1}^{N_{SD,1}}{\tilde{x}_i}$. This implies that the maximum likelihood estimation derived of the superheterodyne model is mathematically equivalent to the arithmetic mean of multiple single-period estimates $\tilde{x}_i$.

\normalem{\emph{Lemma 1:}} The maximum likelihood estimation $\left| \tilde{E}_{ISD} \right|$ under the superheterodyne model, based on $N_{SD,1}$ periods, a total of $N=N_{SD,1}\times N_{SD,2}$ sample points, which averages multiple single-period estimates $\tilde{x}_i$, is mathematically equivalent to performing a discrete fourier transform on the combined multi-period samples $\boldsymbol{z}$ and extracting the amplitude at frequency $\Delta f_s$. The expression is given as follows:
\begin{equation}
	\begin{aligned}
		\left| \tilde{E}_{ISD} \right|&=\frac{1}{N_{SD,1}}\sum_{i=1}^{N_{SD,1}}{\tilde{x}_i}
		\\
		&=\frac{2}{NF_{I}^{\left( 1 \right)}\left[ x_{LO} \right]}\left| \sum_{k=1}^N{z_ke^{-j\frac{2\pi \left( k-1 \right)}{N}N_{SD,1}}} \right|.
	\end{aligned}
\end{equation}

\normalem{\emph{Proof:}} See Appendix \ref{A12}.

\subsubsection{Cramér-Rao Lower Bounds} \label{T323} \textcolor{white}{1}

\normalem{\emph{Theorem 2:}} Under $N$ sampling data and noise $\sigma_{0}$, the CRLB of intensity superheterodyne detection is given as follows,
\begin{equation}
	\begin{aligned}
		\text{CRLB}_{ISD} =\frac{2}{N}\frac{\sigma _{0}^{2} }{\left( F_{I}^{\left( 1 \right)}\left[ x_{LO} \right]\right) ^2 }. 
	\end{aligned}
\end{equation}

\normalem{\emph{Proof:}} Follow the definition of CRLB.

\section{Splitting detection model based on Autler-Townes splitting}\label{T4}

\subsection{Detection Model}\label{T41}
In contrast to intensity detection, the splitting detection methodology necessitates laser frequency scanning $f_p$ to determine the electric field strength $x=\left| E_{RF} \right|$, with principal reliance on the spectral separation between the Autler-Townes splitting peaks rather than absolute APD voltage measurements. Thus, we develop a frequency marginal response function $F_S\left[ f_{p} \right]$ to characterize this measurement modality, given by

\begin{equation}
\begin{aligned}
F_S\left[ f_{p} \right] =\left. \mathcal{G}\left[ x,f_{p} \right] \right|_{x=\left| E_{RF} \right|},
\end{aligned}
\end{equation}
where $F_S\left[ f_{p} \right]$ corresponds to another cross-section of the APD joint response function $\mathcal{G}\left[ x,f_p \right]$ given electric field intensity $x=\left| E_{RF} \right|$. As illustrated in Fig. \ref{diagram}(a), the red curve represents $F_S\left[ f_p \right]$ function at three RF intensity levels.

Without loss of generality, we consider probe laser detuning to complete frequency scanning, and the conclusion is also applicable to the detuning of coupling laser.

The Autler-Townes splitting (defined as $2\pi \Delta f_0$) of the probe laser spectrum is easily measured and under certain conditions is equal to the Rabi frequency of RF transition, given by

\begin{equation}
\begin{aligned}
\text{AT\ splitting}=2\pi \Delta f_0=\Omega _{RF},
\end{aligned}
\end{equation}
where this relationship between the AT splitting and the Rabi frequency $\Omega _{RF}$ is obtained under the weak probe limit and without Doppler averaging. Based on measuring this splitting $\Delta f_m$, the RF strength can be obtained directly. In this approach, either the probe or the coupling laser can be scanned or detuned, given by \cite{holloway2017electric}

\begin{equation}
\begin{aligned}
\left| E_{SF} \right|=\kappa \Delta f_m,
\end{aligned}
\end{equation}
where $\kappa=\frac{2\pi \lambda _p{\hbar }}{\lambda _c\mu _{RF}}$; $\lambda_p$ and $\lambda_c$ are the wavelengths of the probe and coupling laser, respectively; $\Delta f_m=f_{R}-f_{L}$ is the measured splitting that we need for electric field detection, which is obtained by the difference of two peaks $f_{L}$ and $f_{R}$. Figure \ref{diagram}(d) shows the outputs of the splitting detection.

Typically, the two resonance peaks exhibit no correlation in their lineshapes, amplitudes, or relative positions. For instance, under microwave detuning conditions, the peaks demonstrate asymmetric amplitudes. Consequently, estimating the peak splitting is mathematically equivalent to independently determining each peak's frequency shift and computing their difference. Then, the splitting detection problem can be transformed into a peak shift estimation problem. Without loss of generality, we only consider the right peak $f_{R}$ of the two peaks and the estimation of left peak $f_{L}$ can be solved by the same ways. 

A complete frequency scanning process will record two sets of data: the scanning frequency data and scanning voltage data. The collected scanning voltage data, i.e., APD voltage, can be written as a $N_{SF,2} \times N_{SF,1}$ matrix $\boldsymbol{Z}_{R}$, which contains $N_{SF,1}$ scanning frequency points and $N_{SF,2}$ sampling points for each frequency. Matrix $\boldsymbol{Z}_R$ is written as follows,

\begin{equation}
\begin{aligned}
\boldsymbol{Z}_R=\left[ \begin{matrix}
	z_{1,1}&		z_{1,2}&		\cdots&		z_{1,N_{SF,1}}\\
	z_{2,1}&		z_{2,2}&		\cdots&		z_{2,N_{SF,1}}\\
	\vdots&		\vdots&		\ddots&		\vdots\\
	z_{N_{SF,2},1}&		z_{N_{SF,2},2}&		\cdots&		z_{N_{SF,2},N_{SF,1}}\\
\end{matrix} \right] _{N_{SF,2}\times N_{SF,1}}
\end{aligned}
\end{equation}

Then, the $N_{SF,1}$ scanning voltage data $\boldsymbol{z}_{1\times N_{SF,1}}$ for peak estimation is calculated by taking the average for each column of the matrix $\boldsymbol{Z}_{R}$, i.e., $\left[ \boldsymbol{z}_{R} \right] _i=\frac{1}{N_{SF,2}}\sum_{k=1}^{N_{SF,2}}{\left[ \boldsymbol{Z}_{R} \right] _{k,i}}$. At the same time, we also record the laser frequency data and get $N_{SF,1}$ frequency points matrix $\boldsymbol{f}_{R}$. The two sets of scanning data $\boldsymbol{f}_{R}$ and $\boldsymbol{z}_{R}$ are given by
\begin{equation}
\begin{aligned}
\boldsymbol{f}_{R}&=\left[ f_{1},f_{2},\cdots ,f_{N_{SF,1}} \right],
\\
\boldsymbol{z}_{R} &=\left[ z_{1},z_{2},\cdots ,z_{N_{SF,1}} \right],
\end{aligned}
\end{equation}
where we denote $f_i=\left[\boldsymbol{f}_{R} \right]_i $ and $z_i =\left[\boldsymbol{z}_{R} \right]_i$. Since the resonant frequency $f_{p,o}$ of the probe laser-coupled atomic transition remains constant, the receiver can unambiguously distinguish whether the frequency-voltage data $f_i$ and $z_i$ are adopted for the right peak estimation $f_{R}$ by configuring the laser sweep range to be greater than the resonance frequency $f_{p,o}$, i.e., $f_i>f_{p,o}$. In addition, the mathematical framework of the scanning process does not inherently prescribe uniform sampling. Notably, non-uniform sampling strategies yield superior performance in subsequent analyze.

Then, we denote the ideal APD voltage for the $i$-th scanning frequency $f_i$ as $\left[ \boldsymbol{y}_{R} \right] _{i}$ without noise, which is given by
\begin{equation}
\begin{aligned}
\boldsymbol{y}_R=\left[ F_S\left[ f_1 \right] ,F_S\left[ f_2 \right] ,\cdots ,F_S\left[ f_{N_{SF,1}} \right] \right] .
\end{aligned}
\end{equation}

The actual sampling points $\left[ \boldsymbol{Z}_{R} \right] _{k,i}$ follows Gaussian distribution with mean $\left[ \boldsymbol{y}_{R} \right] _{i}$ and variance $\sigma_0 ^2$. Thus, $z_i$, the average of $i$-th column of the matrix $\boldsymbol{Z}_{R}$, follows Gaussian distribution with mean $\left[ \boldsymbol{y}_{R} \right] _{i}$ and variance $\sigma_{SF}^2=\frac{\sigma_0^2}{N_{SF,2}}$. Due to the independence of each scanning frequency point, we have
\begin{equation}
\begin{aligned}
\boldsymbol{z}_R=\boldsymbol{y}_R+\boldsymbol{n}_{SF},
\end{aligned}
\end{equation}
where $\boldsymbol{n}_{SF}$ is $1\times N_{SF,1}$ noise matrix whose elements $\left[ \boldsymbol{n}_{SF} \right] _i$ are statistically independent random variables with identical variance $\mathbb{D}\left[ \left[ \boldsymbol{n}_{SF} \right] _i \right] =\frac{\sigma _{0}^{2}}{N_{SF,2}}$.

Since the frequency marginal response function $F_S\left[ f_{p} \right]$ encapsulates information from both spectral peaks, whereas the estimation of the single peak requires only its localized spectral features. We define $F_{SL}\left[ f \right]$ and $F_{SR}\left[ f \right]$ as the intrinsic lineshape function of the left and right peak, respectively,  given by
\begin{equation}
\begin{aligned}
F_S\left[ f_p \right] =\begin{cases}
	F_{SR}\left[ f_p-f_R \right]&		f_p>f_{p,o}\\
	F_{SL}\left[ f_p-f_L \right]&		f_p<f_{p,o}\\
\end{cases},
\end{aligned}
\end{equation}
where it indicates that the original linear function $F_{SR}\left[ f \right]$ and $F_{SL}\left[ f \right]$ shifted by $f_R$ and $f_L$, respectively, coinciding exactly with the peak function $F_S\left[ f_p \right]$ measured in the experiment.

\subsection{Maximum Likelihood Estimation}\label{T42}

Overall, scanning frequency detection based on Autler-Townes splitting requires solving the inverse problem of estimating peak shift $\tilde{f}_R$ from scanning data $\boldsymbol{f}_{R}$, $\boldsymbol{z}_{R}$. In this work, we categorize the splitting estimation problem into two distinct classes based on the completeness of prior knowledge about the peak function, univariate peak shift estimation, and multivariate peak shift estimation. The former case assumes complete a priori knowledge of the peak function $F_{SR}\left[ f \right]$, including both its lineshape and amplitude—that is, a deterministic peak profile—requiring only the estimation of peak shift $\tilde{f}_R$ from $\boldsymbol{f}_{R}$ and $\boldsymbol{z}_{R}$. The latter case involves partial prior information about the peak function $F_{SR}\left[ \boldsymbol{\theta} \right] =F_{SR}\left[f, \boldsymbol{v} \right] $, necessitating simultaneous estimation of both the peak shift $\tilde{f}_R$ and $M$ lineshape parameters $\boldsymbol{v}=\left[ v_1, v_2,\cdots ,v_M \right] $. As a representative case, a multivariate Gaussian-like peak function $F_{SR}\left[f, \boldsymbol{v} \right] $ can be expressed as
\begin{equation}
\begin{aligned}
F_{SR}\left[f, \boldsymbol{v} \right] =v_1e^{-v_2f^2}+v_3,
\end{aligned}
\end{equation}
where $\boldsymbol{\theta}=\left[ f,\boldsymbol{v}\right] =\left[f,v_1,v_2,v_3 \right] $.

\subsubsection{Univariate Peak Shift Estimation}\label{T421}

Due to the peak lineshape function $F_{SR}\left[ f \right]$ is fully characterized a priori, we only need to estimate the shift parameter $f_R$ based on noisy data $\boldsymbol{z}_R$ and $\boldsymbol{f}_{R}$. Since all samples $\boldsymbol{z}_R=\left[ z_1,z_2,\cdots ,z_{N_{SF,1}} \right] $ are assumed to be independent, the overall likelihood for the entire set of observations is given by
\begin{equation}
\begin{aligned}
&p\left( \boldsymbol{z}_R;f_R \right) =\prod_{i=1}^{N_{SF,1}}{\frac{1}{\sqrt{2\pi \sigma _{SF}^{2}}}}
\\
&\ \ \ \ \ \ \ \ \ \ \ \ \ \ \ \ \ \ \ \times  \exp\left( -\sum_{i=1}^{N_{SF,1}}{\frac{\left( z_i-F_{SR}\left[ f_i-f_R \right] \right) ^2}{2\sigma _{SF}^{2}}}\right) ,
\end{aligned}
\end{equation}
where $\sigma_{SF}^2=\frac{\sigma _{0}^{2}}{N_{SF,2}}$.

The maximum likelihood approach finds the parameter $f_R$ that maximizes the likelihood function $p\left( \boldsymbol{z}_R;f_R \right) $. The peak shift value $\tilde{f}_{UR}$ under univariate estimation can be expressed as

\begin{equation}\label{EqUE}
\begin{aligned}
\tilde{f}_{UR}&=\mathop{\arg\max}\limits_{f_R} \ln p\left( \boldsymbol{z}_R;f_R \right)
\\
&=\mathop{\arg\min}\limits_{f_R} \sum_{i=1}^{N_{SF,1}}{\left( z_i-F_{SR}\left[ f_i-f_R \right] \right) ^2}.
\end{aligned}
\end{equation}

Building upon the solution framework established in work\cite{karl2005high}, we propose an iterative optimization strategy, the complete mathematical derivation is provided in the Appendix \ref{A31}. We begin by initializing parameter $f_{UR}^{\left[ 1 \right]}$, followed by computation of the corresponding $\Delta f_{UR}^{\left[ 1 \right]}$. The iterative solution is then obtained through successive application of the following update formula,
\begin{equation}\label{Ishift}
	\begin{aligned}
		f_{UR}^{\left[ a+1 \right]}=f_{UR}^{\left[ a \right]}+\Delta f_{UR}^{\left[ a \right]},
	\end{aligned}
\end{equation}
where
\begin{equation}\label{Dshift}
	\begin{aligned}
&\Delta f_{UR}^{\left[ a \right]}=
\\
&-\frac{\sum_{i=1}^{N_{SF,1}}{\left( z_i- F_{SR}\left[ f_i-f_{UR}^{\left[ a \right]} \right]  \right) \left( \left. \frac{\partial}{\partial f}F_{SR}\left[ f \right] \right|_{f=f_i-f_{UR}^{\left[ a \right]}} \right)}}{\sum_{i=1}^{N_{SF,1}}{\left( \left. \frac{\partial}{\partial f}F_{SR}\left[ f \right] \right|_{f=f_i-f_{UR}^{\left[ a \right]}} \right) ^2}},
	\end{aligned}
\end{equation}
where the iteration process terminates when $\left| \Delta f_{UR}^{\left[ a \right]}\right| $ falls below a predefined threshold $\varepsilon$, at which point the peak offset value converges to the final solution $\tilde{f}_{UR}=f_{UR}^{\left[ a \right]}$. 


Finally, the electric field strength $\left| \tilde{E}_{UE} \right|$ based on univariate splitting estimation can be regarded as the subtraction of two peaks $\tilde{f}_{UR}$ and $\tilde{f}_{UL}$, given by
\begin{equation}\label{UE}
	\begin{aligned}
		\left| \tilde{E}_{UE} \right|&=\kappa \left( \tilde{f}_{UR}-\tilde{f}_{UL} \right) .
	\end{aligned}
\end{equation}

\subsubsection{Multivariate Peak Shift Estimation}\label{T422}

Due to the peak lineshape function $F_{SR}\left[ f \right]$ is partially prior known, we need to estimate the shift parameter $f_R$ and lineshape parameters $\boldsymbol{v}$ based on noisy data $\boldsymbol{z}_R$ and $\boldsymbol{f}_{R}$. Since all samples $\boldsymbol{z}_R=\left[ z_1,z_2,\cdots ,z_{N_{SF,1}} \right] $ are assumed to be independent, the overall likelihood for the entire set of observations is given by

\begin{equation}
\begin{aligned}
	p\left( \boldsymbol{z}_R; \boldsymbol{\theta} \right) &= \prod_{i=1}^{N_{SF,1}}{\frac{1}{\sqrt{2\pi \sigma _{SF}^{2}}}}
\\
&\ \ \ \times \exp \left( -\sum_{i=1}^{N_1}{\frac{\left( z_i-F_{SR}\left[ f_i-f_R,\boldsymbol{v} \right] \right) ^2}{2\sigma _{SF}^{2}}} \right)  .
\end{aligned}
\end{equation}

Then, the maximum likelihood approach finds the parameter $f_R$ and $\boldsymbol{v}$ that maximizes the likelihood function $p\left( \boldsymbol{z}_R;\left[f_R, \boldsymbol{v}\right]  \right) $. The peak shift value $\tilde{f}_{MR}$ under multivariate estimation can be expressed as

\begin{equation}\label{MP}
\begin{aligned}
\tilde{f}_{MR}&=\mathop{\arg\max}\limits_{\boldsymbol{v},f_R} \ln p\left( \boldsymbol{z}_R;\left[f_R, \boldsymbol{v}\right]  \right)
\\
&=\mathop{\arg\min}\limits_{\boldsymbol{v},f_R} \sum_{i=1}^{N_{SF,1}}{\left( z_i-F_{SR}\left[ f_i-f_R,\boldsymbol{v} \right] \right) ^2}.
\end{aligned}
\end{equation}

Here, we give an alternating iterative optimization strategy to solve Eq. (\ref{MP}) and other classical iterative methods can also obtain numerical results. We commence the optimization procedure by initializing both $f_{MR}^{\left[ 1\right] }$ and $\boldsymbol{v}^{\left[ 1\right] }$. The algorithm then proceeds through alternating optimization phases: 

\begin{itemize}
	\item[$\bullet$] Phase 1: With $\boldsymbol{v}^{\left[ a\right] }$ held constant, $f_{MR}^{\left[ a+1\right] }$ is iteratively computed using Eq. (\ref{Ishift}) and Eq. (\ref{Dshift}) until convergence.
\end{itemize}
\begin{itemize}
	\item[$\bullet$] Phase 2: With $f_{MR}^{\left[ a+1\right] }$  obtained from the Phase 1 held constant, $\boldsymbol{v}^{\left[ a+1\right] }$ is optimized via BFGS iterative method until convergence.
\end{itemize}

Similarly, the electric field strength $\left| \tilde{E}_{ME} \right|$ based on univariate splitting estimation can be regarded as the subtraction of two peaks $\tilde{f}_{MR}$ and $\tilde{f}_{ML}$, which is given by
\begin{equation}\label{ME}
	\begin{aligned}
		\left| \tilde{E}_{ME} \right|&=\kappa \left( \tilde{f}_{MR}-\tilde{f}_{ML} \right) .
	\end{aligned}
\end{equation}

Overall, in order to obtain one RF field strength at a certain time, the required sampling points are $N=2N_{SF,1}\times N_{SF,2}$ and at least $2N_{SF,1}$ due to frequency scanning.

\subsection{Cramér-Rao Lower Bounds} \label{T43} 
\subsubsection{CRLB of Univariate Peak Shift Estimation}\label{T431}\textcolor{white}{1}

\normalem{\emph{Theorem 3:}} Under $N=2N_{SF,1}\times N_{SF,2}$ sampling data and noise $\sigma_{0}$, the CRLB of univariate splitting detection $\left| \tilde{E}_{UE} \right|$ is given as follows,

\begin{equation}
	\begin{aligned}
\text{CRLB}_{UE}&=\kappa ^2\left( \text{CRLB}_{UR}+\text{CRLB}_{UL} \right) 
\\
&= \frac{\kappa ^2\sigma _{0}^{2}}{N_{SF,2}\sum_{i=1}^{N_{SF,1}}{\left( F_{SR}^{\left( 1 \right)}\left[ f_i-f_R \right] \right) ^2}} 
\\
&\ \ \ + \frac{\kappa ^2\sigma _{0}^{2}}{N_{SF,2}\sum_{i=1}^{N_{SF,1}}{\left( F_{SL}^{\left( 1 \right)}\left[ f_i-f_L \right] \right) ^2}} ,
	\end{aligned}
\end{equation}
where this analysis results are based on the estimation and measurement that the two peaks are independent of each other.  $\text{CRLB}_{UR}$ and $\text{CRLB}_{UL}$ are the CRLBs for the two peak shifts.

\normalem{\emph{Proof:}} See Appendix \ref{A32}.

\subsubsection{CRLB of Multivariate Peak Shift Estimation}\label{T433}\textcolor{white}{1}

\normalem{\emph{Theorem 4:}} Under $N=2N_{SF,1}\times N_{SF,2}$ sampling data and noise $\sigma_{0}$, the CRLB of multivariable splitting detection $\left| \tilde{E}_{ME} \right|$ is given as follows,

\begin{equation}
	\begin{aligned}
		\text{CRLB}_{ME}&=\kappa^2\left( \text{CRLB}_{MR}+\text{CRLB}_{ML}\right) ,
	\end{aligned}
\end{equation}
where the analysis results are also based on the independence of two peaks. $\text{CRLB}_{MR}$ and $\text{CRLB}_{ML}$ are the CRLBs of multivariable estimation for the two peak shifts.

\normalem{\emph{Proof:}} See Appendix \ref{A33}.

\section{Comparison and analysis of detection methods}\label{T5}

%
%

In our comparative analysis, we employ the CRLB as the benchmark to evaluate the performance of different detection schemes.

\subsection{Intensity Direct Detection vs Intensity Superheterodyne Detection}\label{T51}


In the direct detection scheme, the received APD voltage $z$ is converted to the corresponding the measured electric field $x$ through the inverse function of $F_I\left[ x \right]$. Conversely, the superheterodyne detection approach bypasses this inverse transformation, instead exploiting the approximately linear relationship between the APD's difference-frequency component amplitude and the electric field strength $x$ under weak signal conditions. Such fundamental distinction results in a critical difference in their noise characteristics: for direct detection, the equivalent noise variance is $\frac{\sigma _{0}^{2}}{\left( F_I^{\left( 1 \right)}\left[ x \right] \right) ^2}$ depends functionally on the estimated variable $x$, whereas for superheterodyne detection, the equivalent noise variance $\frac{\sigma _{ISD}^{2}}{N_{SD,1}}=\frac{2\sigma _{0}^{2}}{N\left( F_I^{\left( 1 \right)}\left[ x_{LO} \right] \right) ^2}$ is determined solely by the local oscillator intensity $x_{LO}$ and remains independent of the estimated variable $x$.


Assuming both detection schemes acquire $N$ total sampling points, their respective lower bounds on estimation variance are given by
\begin{equation}
\begin{aligned}
\mathbb{D}\left[ \left| \tilde{E}_{IDD} \right| \right] &\ge \text{CRLB}_{IDD}=\frac{\sigma _{0}^{2}}{N\left( F_{I}^{\left( 1 \right)}\left[ x \right] \right) ^2},			
\\
\mathbb{D}\left[ \left| \tilde{E}_{ISD} \right| \right] &\ge \text{CRLB}_{ISD}= \frac{2\sigma _{0}^{2}}{N\left( F_{I}^{\left( 1 \right)}\left[ x_{LO} \right] \right) ^2}
\\
&\overset{1}{\ge}\frac{2\sigma _{0}^{2}}{N\max_{x_{LO}}\left( F_{I}^{\left( 1 \right)}\left[ x_{LO} \right] \right) ^2},
\end{aligned}
\end{equation}
where $\overset{1}{\ge}$ demonstrates that the variance decreases with increasing slope magnitude, with the minimum corresponding to the maximum slope configuration.

Overall, there are two fundamental mechanisms causing the gap in their ability to detect signals with lower power. Firstly, for intensity direct detection, the noise is related to the estimated signal strength $x$ due to factor $\frac{1}{F_{I}^{\left( 1 \right)}\left[ x \right]}$. Conversion function $F\left[ x \right]$ usually has a smaller slope for signals with lower power, which finally causes significant noise and measurement uncertainty. On the other hand, for moderate power signal measurements where the measured electric field strength $x$ exhibits a sufficiently large slope $F_{I}^{\left( 1 \right)}\left[ x \right]$, intensity direct detection can achieve superior performance. For superheterodyne detection, the noise is related to local oscillator signal $x_{LO}$ rather than estimated signal strength $x$. Therefore, adjusting the working point appropriately to make the system have a smaller $\frac{1}{F^{\left( 1 \right)}\left[ x_{LO} \right]}$ can improve sensitivity for low-power signals. The second factor arises from the estimation methodology. Although practical receiver systems may encounter more complex Gaussian noise profiles (e.g., signal-dependent Gaussian noise), the superheterodyne architecture fundamentally maintains noise characteristics that remain uncorrelated with the estimated variables $x$. Such unique property enables direct extraction of the maximum likelihood estimate of the electric field strength $x$ through simple discrete Fourier transform processing. In contrast, for intensity direct detection, the presence of signal-correlated Gaussian noise implies that simply averaging APD voltage readings $\tilde{x}=F_{I}^{-1}\left[ \mathbb{E}\left[ \boldsymbol{z} \right] \right] $ does not yield the maximum likelihood estimate of the electric field strength $x$.

\normalem{\emph{Observation 1:}} The detection performance of intensity detection depends on the slope $F_I^{\left(1 \right) }\left[ x \right] $ of the intensity marginal response function $F_I\left[ x \right] $:
\begin{itemize}
	\item[$\bullet$] Intensity direct detection can achieve superior performance when the measured electric field strength $x$ exhibits a sufficiently large slope for $\left| F_{I}^{\left( 1 \right)}\left[ x \right]\right| $.
\end{itemize}
\begin{itemize}
	\item[$\bullet$] Intensity superheterodyne detection can achieve superior performance by adjusting the working point $x_{LO}$ to exhibit a sufficiently large slope for $\left| F_{I}^{\left( 1 \right)}\left[ x_{LO} \right]\right| $.
\end{itemize}

\subsection{Univariate Peak Shift Estimation Vs Multivariate Peak Shift Estimation}\label{T52}


For an $M+1$ dimensional parameter vector $\boldsymbol{\theta }$ and its Fisher information matrix $\mathbf{J}_{M+1}\left[\boldsymbol{\theta }  \right]$, the CRLB is given by $\mathbf{J}_{M+1}^{-1}$. When partial prior knowledge is available, let $\boldsymbol{\theta}'$ denote the $M$ dimensional parameter vector, We have $\boldsymbol{\theta}=\left[\boldsymbol{\theta}',\theta_{M+1} \right] $. The Fisher information matrix and CRLB are $\mathbf{J}_M\left[\boldsymbol{\theta }'\right]$ and $\mathbf{J}_{M}^{-1}$, respectively. Using block matrix inversion, we have
\begin{equation}
\begin{aligned}
\mathbf{J}_{M+1}^{-1}&=\left[ \begin{matrix}
	\mathbf{J}_M&		\boldsymbol{b}\\
	\boldsymbol{b}^T&		c\\
\end{matrix} \right] ^{-1}=\left[ \begin{matrix}
	\mathbf{J}_{M}^{-1}+\mathbf{J}_{M}^{-1}\boldsymbol{b}\tilde{c}\boldsymbol{b}^T\mathbf{J}_{M}^{-1}&		-\mathbf{J}_{M}^{-1}\boldsymbol{b}\tilde{c}\\
	-\tilde{c}\boldsymbol{b}^T\mathbf{J}_{M}^{-1}&		\tilde{c}\\
\end{matrix} \right] ,
\end{aligned}
\end{equation}
where $\boldsymbol{b}$ and its transpose $\boldsymbol{b}^T$ are vector. $c$ is constant. Since the Fisher information matrix is positive definite, it follows that $\tilde{c}=\left( c-\boldsymbol{b}^T\mathbf{J}_{M}^{-1}\boldsymbol{b} \right) ^{-1}>0$ and $\mathbf{J}_{M}^{-1}\boldsymbol{b}\tilde{c}\boldsymbol{b}^T\mathbf{J}_{M}^{-1}$ constitutes a positive semi-definite matrix. For the diagonal elements, we obtain $\left[ \mathbf{J}_{M+1}^{-1} \right] _{m,m}\ge \left[ \mathbf{J}_{M}^{-1} \right] _{m,m}$, where $m=1,2,\cdots,M$. The CRLB analysis reveals that increasing the number of estimated parameters (or equivalently, reducing prior knowledge of the peak function) inherently degrades the estimation precision of individual peak displacements, thereby compromising the measurement accuracy of peak splitting. Therefore, the optimal estimation scheme for splitting detection requires complete prior knowledge of the peak function, with only the peak shift treated as an unknown parameter, i.e., univariate peak shift estimation.

Based on the CRLB, we derive the minimum variance for estimating the electric field strength $\left| \tilde{E}_{UE} \right|$ using univariate peak shift estimation, given by

\begin{equation}
\begin{aligned}
&\mathbb{D}\left[ \left| \tilde{E}_{UE} \right| \right] =\kappa ^2\mathbb{D}\left[ \tilde{f}_{UR} \right] +\kappa ^2\mathbb{D}\left[ \tilde{f}_{UL} \right] 
\\
&\ge \kappa ^2\frac{1}{-\mathbb{E}\left\{ \frac{\partial ^2}{\partial f_{R}^{2}}p\left( \boldsymbol{z}_R;f_R \right) \right\}}+\kappa ^2\frac{1}{-\mathbb{E}\left\{ \frac{\partial ^2}{\partial f_{L}^{2}}p\left( \boldsymbol{z}_L;f_L \right) \right\}}
\\
&=\kappa ^2\frac{\sigma _{0}^{2}}{N_{SF,2}\sum_{i_R=1}^{N_{SF,1}}{\left( F_{SR}^{\left( 1 \right)}\left[ f_{i_R}-f_R \right] \right) ^2}}
\\
&\ \ \ \ \ +\kappa ^2\frac{\sigma _{0}^{2}}{N_{SF,2}\sum_{i_L=1}^{N_{SF,1}}{\left( F_{SL}^{\left( 1 \right)}\left[ f_{i_L}-f_R \right] \right) ^2}}
\\
&\overset{1}{\ge} \frac{\kappa ^2\sigma _{0}^{2}}{N_{SF,1}N_{SF,2}\max_{i_R}\left( F_{SR}^{\left( 1 \right)}\left[ f_{i_R}-f_R \right] \right) ^2}
\\
&\ \ \ \ \ +\frac{\kappa ^2\sigma _{0}^{2}}{N_{SF,1}N_{SF,2}\max_{i_L}\left( F_{SL}^{\left( 1 \right)}\left[ f_{i_L}-f_R \right] \right) ^2}
\\
&\overset{2}{\ge} \frac{4\kappa ^2\sigma _{0}^{2}}{N\max_{f_i}\left( F_{S}^{\left( 1 \right)}\left[ f_i \right] \right) ^2},
\end{aligned}
\end{equation}
where in $\overset{1}{\ge}$, since the CRLB for peak shift depends on the slopes of both peak functions, which appear squared in the denominator of the CRLB expression, the variance bound is necessarily larger than that obtained when all sampling points are concentrated at the maximum slope regions. In $\overset{2}{\ge}$, we have $N=2N_{SF,1}\times N_{SF,2}$. The frequency marginal response function $F_S\left[ f_{p} \right]$ is composed of two peak functions $F_{SL}\left[ f_p-f_L \right]$ and $F_{SR}\left[ f_p-f_R \right]$, leading to the mathematical result that the sum of the reciprocals of the individual peak's maximum slopes exceeds the reciprocal of the edge response function's maximum slope. However, from a physical implementation perspective, this does not imply that single-point sampling at the maximum slope of the $F_S\left[ f_{p} \right]$ is sufficient. Rather, it necessitates independent sampling at the respective maximum-slope regions of both constituent peak functions.

\normalem{\emph{Observation 2:}} The detection performance of splitting detection depends on the slope $F_S^{\left(1 \right) }\left[ f_{p} \right] $ and the prior knowledge of the frequency marginal response function $F_S\left[ f_{p} \right]$:
\begin{itemize}
\item[$\bullet$] Sampling preferentially at the maximum-slope regions of the frequency marginal response function $F_S\left[ f_{p} \right]$ can achieve superior performance.
\end{itemize}
\begin{itemize}
\item[$\bullet$] Incorporating complete prior knowledge of the $F_S\left[ f_{p} \right]$ function's characteristics, including its spectral shape and profile, can achieve superior performance.
\end{itemize}

\subsection{Intensity Detection Vs Splitting Detection}\label{T53}


For a given weak electric field strength $x$ and lasers parameters, the ratio of the minimum variances between intensity superheterodyne detection and splitting detection is given by
\begin{equation}
\begin{aligned}
\frac{\min \mathbb{D}\left[ \left| \tilde{E}_{UE} \right| \right]}{\min \mathbb{D}\left[ \left| \tilde{E}_{ISD} \right| \right]}&=\frac{\frac{4\kappa ^2\sigma _{0}^{2}}{N\max_{f_i}\left( F_{S}^{\left( 1 \right)}\left[ f_i \right] \right) ^2}}{\frac{2\sigma _{0}^{2}}{N\max_{x_{LO}}\left( F_{I}^{\left( 1 \right)}\left[ x_{LO} \right] \right) ^2}}
\\
&=2\kappa ^2\frac{\max_{x_{LO}}\left( F_{I}^{\left( 1 \right)}\left[ x_{LO} \right] \right) ^2}{\max_{f_i}\left( F_{S}^{\left( 1 \right)}\left[ f_i \right] \right) ^2}
\\
&=\left( \frac{\max_{x_{LO}}\left| \left. \frac{\partial \mathcal{G}\left[ \sqrt{2}\kappa x,f_p \right]}{\partial x} \right|_{x=x_{LO}} \right|}{\max_{f_i}\left| \left. \frac{\partial \mathcal{G}\left[ \sqrt{2}\kappa x,f_p \right]}{\partial f_p} \right|_{f_p=f_i} \right|} \right) ^2 \triangleq r_0^2,
\end{aligned}
\end{equation}
where the ratio $r_0$ reflects the intrinsic properties of the APD joint response function $\mathcal{G}\left[ x,f_p \right]$, which can be seen as an indicator of the basic detection capability of two schemes. When $r_0>1$, it conclusively demonstrates that the superheterodyne detection outperforms splitting detection. 

For more general cases, given an electric field strength $x$ and lasers parameters, the ratio between the minimum variances of splitting detection and direct detection is given by
\begin{equation}
\begin{aligned}
\frac{\min \mathbb{D}\left[ \left| \tilde{E}_{UE} \right| \right]}{\mathbb{D}\left[ \left| \tilde{E}_{IDD} \right| \right]}&=\frac{\frac{4\kappa ^2\sigma _{0}^{2}}{N\max_{f_i}\left( F_{S}^{\left( 1 \right)}\left[ f_i \right] \right) ^2}}{\frac{\sigma _{0}^{2}}{N\left( F_{I}^{\left( 1 \right)}\left[ x \right] \right) ^2}}
\\
&=\left( \frac{\left| \left. \frac{\partial \mathcal{G}\left[ 2\kappa x,f_p \right]}{\partial x} \right|_{x=x} \right|}{\max_{f_i}\left| \left. \frac{\partial \mathcal{G}\left[ 2\kappa x,f_p \right]}{\partial f_p} \right|_{f_p=f_i} \right|} \right) ^2 \triangleq r\left[x \right]^2,
\end{aligned}
\end{equation}
where $r\left[x \right]$ serves as a quantitative figure of merit to evaluate and compare the electric field measurement capabilities of two schemes for a given field strength $x$.

Furthermore, it is noteworthy that the coefficient $r_0$ remains constant, whereas $r\left[x \right]$ is a function of the electric field strength $x$ for given lasers configurations. This characteristic stems from the inherent requirement of superheterodyne detection that the measured signal $x$ must remain relatively weak. Consequently, ratio $r_0$ can be regarded as a quantitative metric comparing the weak-signal detection capabilities of the superheterodyne detection and splitting detection. 


\normalem{\emph{Observation 3:}} The detection performance of a given Rydberg system $\mathcal{G}\left[ x,f_p \right]$ for specified electric field strength $x$ depends on the slope $\frac{\partial \mathcal{G}\left[ x,f_p \right]}{\partial x}$ and $\frac{\partial \mathcal{G}\left[ x,f_p \right]}{\partial f_p}$ of the joint response function $\mathcal{G}\left[ x,f_p \right]$:

\begin{itemize}
	\item[$\bullet$] Intensity superheterodyne detection outperforms splitting detection when $r_0 >1$ for the weak measured electric field strength $x$.
\end{itemize}
\begin{itemize}
	\item[$\bullet$] Intensity direct detection outperforms splitting detection when $r\left[ x\right] >1$ for the measured electric field strength $x$.
\end{itemize}

\section{NUMERICAL RESULTS}\label{T6}
Our numerical simulations employ the Rubidium atomic system illustrated in Fig. \ref{diagram}(a) and its corresponding normalized joint response function $\mathcal{G}\left[ x,f_{p} \right]$ shown in Fig. \ref{diagram}(b), where $\mathcal{G}\left[ 0,f_{p,o} \right]= 1$ serves as the reference point.

Crucially, the linear relationship between electric field strength $x=\left|E_{RF} \right| $ and Rabi frequency $\Omega_{RF}$ established in Eq. (\ref{OE}) enables direct equivalence between electric field estimation $\left|\tilde{E}_{RF} \right| $ and Rabi frequency estimation $\tilde{\Omega}_{RF} $. We have 

\begin{equation}
	\begin{aligned}
		\mathbb{D}\left[ \tilde{\Omega}_{RF} \right] =\frac{\mu _{RF}^{2}}{{\hbar }^2}\mathbb{D}\left[ \left| \tilde{E}_{RF} \right| \right] ,
	\end{aligned}
\end{equation}
where $\frac{\mu _{RF}^{2}}{{\hbar }^2}$ is a constant. Consequently, the estimation of electric field strength $\left|\tilde{E}_{RF} \right| $ becomes mathematically equivalent to Rabi frequency estimation $\tilde{\Omega}_{RF} $, differing only by a constant proportionality factor $\frac{\mu _{RF}^{2}}{{\hbar }^2}$ that is determined by the atomic dipole moment. Throughout this work, all presented CRLB and mean squared error (MSE) correspond specifically to Rabi frequency estimation, expressed in $\frac{1}{2\pi}$ MHz for direct physical interpretation.

\subsection{Iterative Convergence and Simulation Results of Splitting Detection}

\begin{figure}[htbp]
	\centering
	\includegraphics[width=0.45\textwidth]{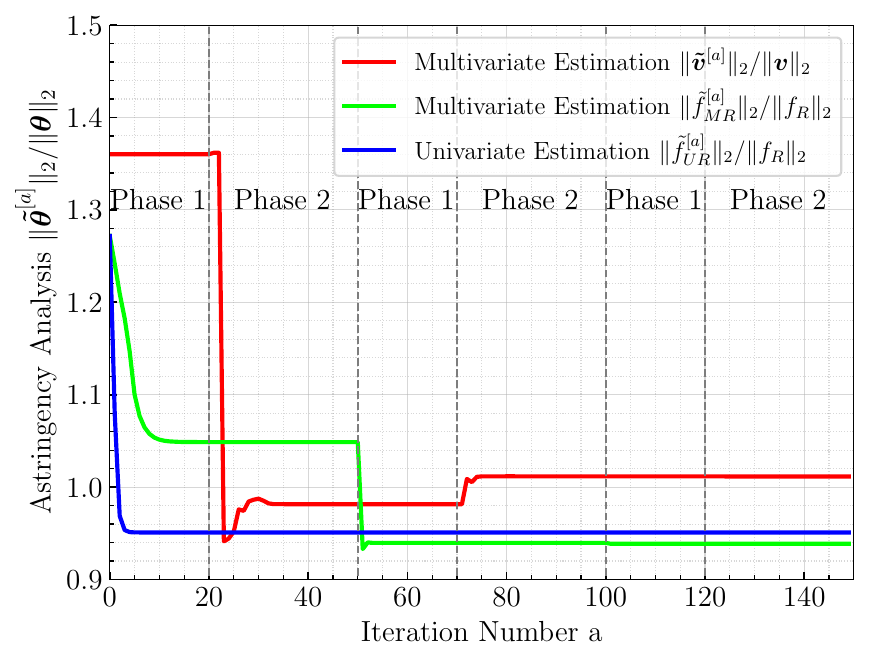}
	\caption{The iterative process of univariate and multivariate peak shift estimation.}
	\label{PeakIter}
\end{figure}

\begin{figure}[htbp]
	\centering
	\includegraphics[width=0.45\textwidth]{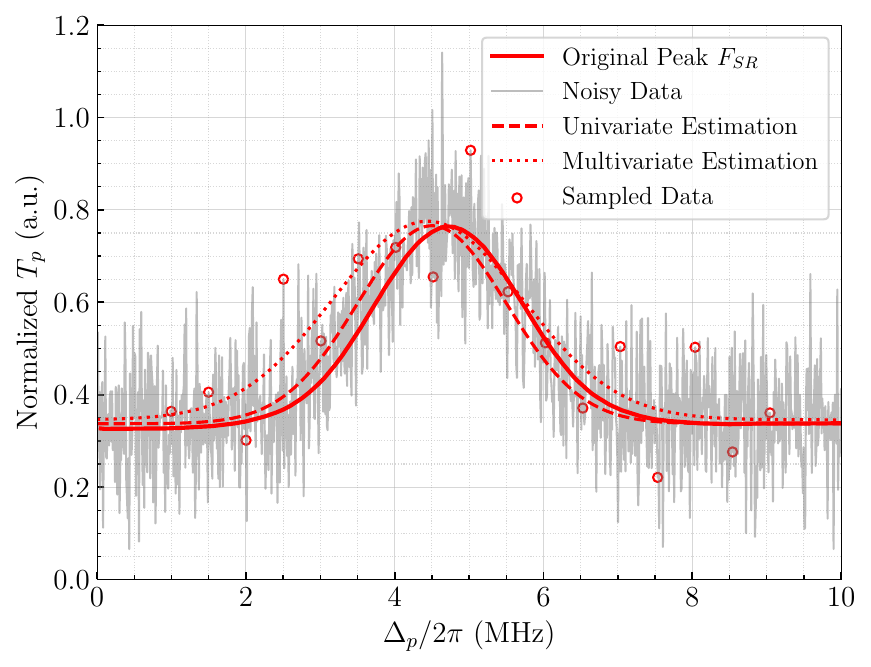}
	\caption{Single right peak shift estimation result, $\Omega_{RF}/2\pi=15$ MHz and $\sigma_{0}=0.1$.}
	\label{PeakEstimation}
\end{figure}

Figures \ref{PeakIter} and \ref{PeakEstimation} present the iterative optimization process and final estimation results, respectively. In Fig. \ref{PeakIter}, we illustrate the convergence behavior of our algorithm through successive iterations. The vertical axis represents the ratio of the L2-norm between the estimated values $\lVert \boldsymbol{\tilde{\theta}}^{\left[ a \right]} \rVert _2$ from $a$-th iteration and the ground truth values $\lVert \boldsymbol{\theta } \rVert _2$. Figure \ref{PeakEstimation} demonstrates the estimation results for the spectral peak parameters under 17 sampled noisy data.

Furthermore, for splitting detection, we deliberately neglect the inherent nonlinear relationship between Autler-Townes splitting and applied RF field strength under weak-signal conditions \cite{holloway2017electric}, instead adopting a simplified linear approximation across all RF power regimes. 

In Tables \ref{combined_table}, we present a comparative analysis of the detection variance and the corresponding CRLB under two distinct sampling strategies: averaged sampling and maximum-slope sampling, both evaluated with $N = 20$ sampling points. We demonstrate the averaged sampling strategy, where 10 measurement points are uniformly distributed across the $\Delta_p/2\pi=10$ MHz frequency sweep range (yielding 20 total samples for the double peaks), providing baseline performance for conventional detection approaches. In contrast, Table \ref{combined_table} implements the maximum-slope sampling strategy derived in Section \ref{T42}, concentrating all 20 sampling points around the maximum slope of the spectral region. We also include the conventional polynomial fitting results for peak shift estimation, which adopts fifth-order polynomial function denoted as 5-PF. Building upon the maximum likelihood estimation framework established in Section. \ref{T4}, we evaluate the estimation performance through mean squared error analysis for both univariate estimation (UE) and multivariate estimation (ME) approaches, along with their corresponding theoretical limits given by the CRLB (U-CRLB for univariate and M-CRLB for multivariate cases). The numerical results conclusively validate the two fundamental principles established in Section. \ref{T52}: (1) optimal sampling at maximum-slope regions and (2) incorporation of complete a priori peak function knowledge. Our maximum-likelihood shift estimation method achieves superior performance compared to conventional polynomial fitting.

%
%
%
%
%
%
\begin{table*}
\caption{The MSE and CRLB of splitting detection under different sampling methods ($\frac{1}{2\pi}\times 10^{-3}$ MHz) \label{combined_table}}
\centering
\begin{tabular}{|c|c|c|c|c|c|c|c|c|}
	\hline
	Sampling Method & Siganl Intensity & Noise Variance & UE & ME & 5-PF & U-CRLB & M-CRLB \\
	\hline
	\multirow{2}{*}{Average Sampling} 
	& $\Omega_{RF}/2\pi=15$ MHz & $\sigma_{0}=0.01$ & 3.8243 & 3.8282 & 12.2174 & 3.8006 & 3.8038 \\
	\cline{2-8}
	& $\Omega_{RF}/2\pi=15$ MHz & $\sigma_{0}=0.02$ & 15.3672 & 15.472 & 48.6678 & 15.2024 & 15.2152 \\
	\hline
	\multirow{2}{*}{Maximum Slope Sampling} 
	& $\Omega_{RF}/2\pi=15$ MHz & $\sigma_{0}=0.01$ & 0.9051 & 0.9358 & 51.6937 & 0.8739 & 0.8739 \\
	\cline{2-8}
	& $\Omega_{RF}/2\pi=15$ MHz & $\sigma_{0}=0.02$ & 3.3946 & 3.7613 & 251.7924 & 3.4956 & 3.4956 \\
	\hline
	\multirow{2}{*}{Average Sampling} 
	& $\Omega_{RF}/2\pi=10$ MHz & $\sigma_{0}=0.01$ & 4.0540 & 4.0643 & 5.2852 & 3.9368 & 3.9401 \\
	\cline{2-8}
	& $\Omega_{RF}/2\pi=10$ MHz & $\sigma_{0}=0.02$ & 16.2185 & 16.342 & 21.3313 & 15.7472 & 15.7605 \\
	\hline
	\multirow{2}{*}{Maximum Slope Sampling} 
	& $\Omega_{RF}/2\pi=10$ MHz & $\sigma_{0}=0.01$ & 0.9024 & 0.9307 & 51.5614 & 0.8913 & 0.8913 \\
	\cline{2-8}
	& $\Omega_{RF}/2\pi=10$ MHz & $\sigma_{0}=0.02$ & 3.6260 & 4.0497 & 240.8377 & 3.5652 & 3.5652 \\
	\hline
\end{tabular}
\end{table*}

\subsection{Simulation Results of Comparison}

As derived in Section. \ref{T5}, the CRLB for all detection schemes depends on three key parameters: (1) the number of sampling points $N$, (2) the noise variance $\sigma_{0}^2$, and (3) the slope of $\mathcal{G}\left[ x,f_{p} \right]$. To enable unified comparison across different experimental configurations, we normalize both the CRLB and MSE results by dividing them by $\frac{\sigma _{0}^{2}}{N\max_{x_{LO}}\left( F_{I}^{\left( 1 \right)}\left[ x_{LO} \right] \right) ^2}$. 

\begin{figure}[htbp]
	\centering
	\includegraphics[width=0.45\textwidth]{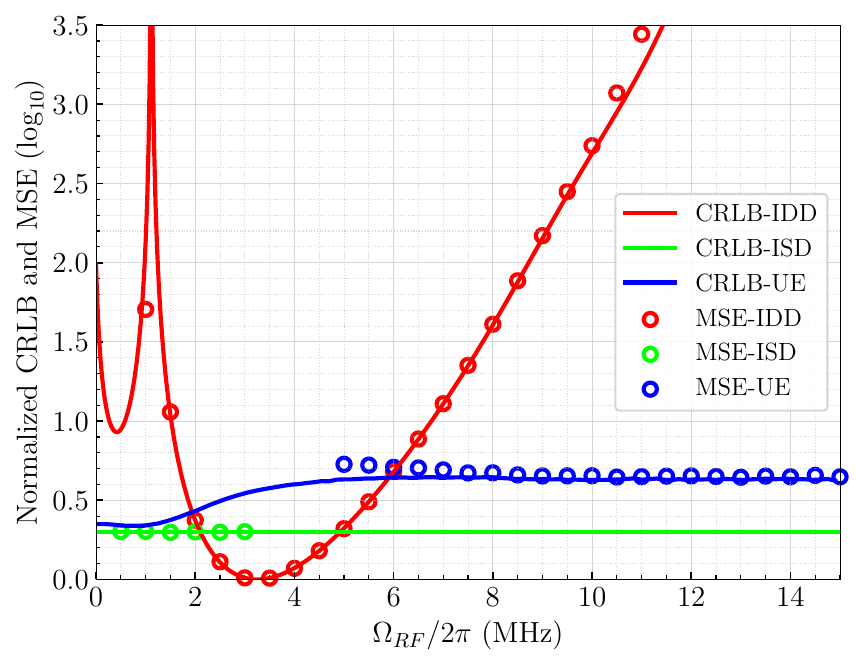}
	\caption{The normalized CRLB and MSE for different detection schemes, $\sigma_{0}=0.01$ and $N=20$.}
	\label{CRLBMSE1}
\end{figure}

\begin{figure}[htbp]
	\centering
	\includegraphics[width=0.45\textwidth]{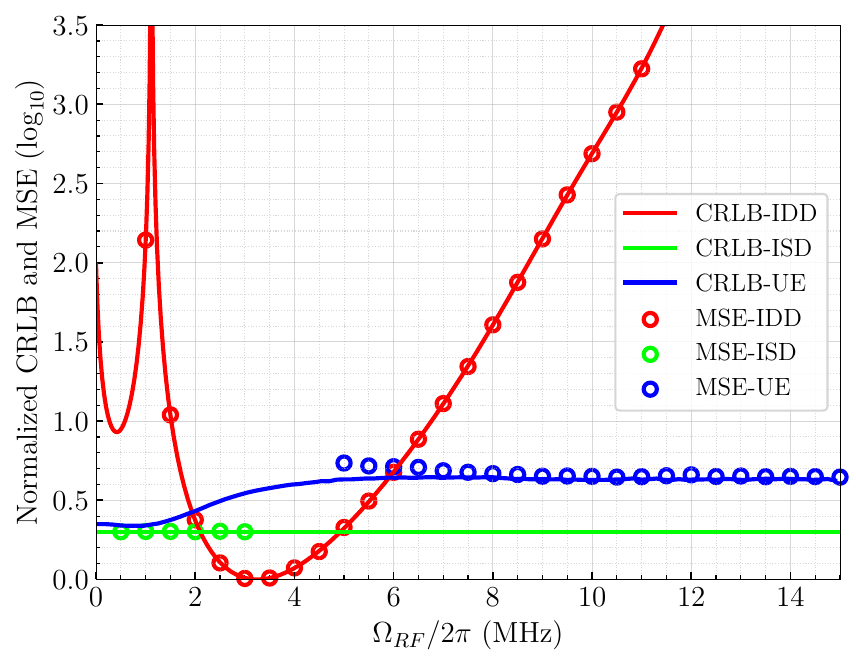}
	\caption{The normalized CRLB and MSE for different detection schemes, $\sigma_{0}=0.001$ and $N=40$.}
	\label{CRLBMSE2}
\end{figure}

In Fig. \ref{CRLBMSE1} and Fig. \ref{CRLBMSE2}, we present a comprehensive comparison of the detection performance across various measurement schemes as a function of the parameter intensity $\Omega_{RF}/2\pi$, with results quantified through normalized CRLB and MSE. Although different noise levels $\sigma_{0}$ and sampling points $N$ are configured in these two figures, the normalized results show no significant differences. This confirms the conclusion demonstrated in Section. \ref{T5}, that the optimal detection performance is determined by the joint response function $\mathcal{G}\left[ x,f_p \right]$, an intrinsic property of the given Rydberg system. The normalized results in this two figures provide a rigorous per-sample efficiency comparison, revealing fundamental differences in detection scheme effectiveness under unit sampling conditions. 

As concluded in Section. \ref{T5}, the detection performance of a given Rydberg system $\mathcal{G}\left[ x,f_p \right]$ for specified electric field strength $x$ depends on the slope $\frac{\partial \mathcal{G}\left[ x,f_p \right]}{\partial x}$ and $\frac{\partial \mathcal{G}\left[ x,f_p \right]}{\partial f_p}$ of the joint response function $\mathcal{G}\left[ x,f_p \right]$. The CRLB for intensity direct detection depends on the slope of $F_I\left[x \right] $, as indicated by the black curve in Fig. \ref{diagram}(a). Except for the region where the slope gradually flattens as $\Omega_{RF}$ approaches $2\pi \times 1$ MHz, the overall slope initially increases and then decreases. Thus, as $\Omega_{RF}/2\pi$ approaches $1$ MHz, the CRLB shows an increasing trend. For $\Omega_{RF}/2\pi$ between $2\sim 6$ MHz, steeper slope enables the Rydberg system to achieve superior detection performance for signal strengths within this range. Beyond 6 MHz, the progressively decreasing slope degrades the detection performance, falling below that of splitting detection. For weak electric field signals, intensity superheterodyne detection performance depends on the operating point of the local oscillator electric field $x_{LO}$. Its optimal CRLB equals twice the minimum CRLB of intensity direct detection. The CRLB for splitting detection depends on the slope of $F_S\left[f_p \right] $, as indicated by the red curve in Fig. \ref{diagram}(a). When $\Omega_{RF}/2\pi$ exceeds 5 MHz, the two split peaks become fully separated. Their spectral shapes remain independent of the signal intensity, with only the peak positions shifting. Consequently, the CRLB remains stable in this regime. In contrast, when $\Omega_{RF}/2\pi$ falls below 5 MHz, the two peaks overlap significantly. This overlap increases the slope of the merged peak, leading to a reduction in the CRLB.

Notably, for the atomic system presented in this work, the parameter $r_0$ is always larger than 1, as shown in Fig. \ref{CRLBMSE1}. This implies that even if the splitting detection maintains a linear response in the weak-signal regime, its detection capability remains inferior to that of superheterodyne detection. In contrast, $r\left[x \right]$ less than 1 within a specific interval $\Omega_{RF}/2\pi >6$ MHz, indicating that the splitting detection, based on univariate peak shift estimation, outperforms that of direct intensity detection in this range. The MSE results confirm our analysis conclusion.

Finally, although the results in this work are based on the classic four-level system, our theoretical framework provides a foundational tool for benchmarking emerging Rydberg sensor architectures. Given the universal nature of this limitation in Rydberg-based perception, the analytical conclusions are equally applicable to alternative configurations, including optical homodyne structures\cite{schlossberger2024rydberg}.

\section{Conclusion} \label{T7}
In this work, we have systematically categorized and modeled existing signal readout methods, classifying them into these two paradigms. We have derived the maximum likelihood estimation procedures and corresponding Cramér-Rao lower bounds for each detection model. Through the analysis of CRLB, we propose the strategies to minimize the estimated variance and improve sensitivity for two detection schemes. The comparative analysis also reveals the optimal detection performance of the two detection schemes. Numerical results demonstrate the effectiveness of our proposed maximum likelihood estimation methods.

\section{Acknowledgement}
The authors would like to thank Mr. Xie Chongwu for discussions.

\appendices

\section{MLE of Intensity Superheterodyne Detection}\label{A12}

The maximum likelihood estimation based on $N_{SD,1}$ periods, with a total of $N=N_{SD,1}\times N_{SD,2}$ sample points, can be written as 
\begin{equation}
	\begin{aligned}
		\left| \tilde{E}_{ISD} \right|&=\frac{1}{N_{SD,1}}\sum_{i=1}^{N_{SD,1}}{\tilde{x}_i}
		\\
		&=\frac{2}{F_{I}^{\left( 1 \right)}\left[ x_{LO} \right] N_{SD,2}N_{SD,1}}
		\\
		&\ \ \ \times \sum_{i=1}^{N_{SD,1}}{\left| \sum_{k=1}^{N_{SD,2}}{z_{k+\left( i-1 \right) \times N_{SD,2}}e^{-j\frac{2\pi \left( k-1 \right)}{N_{SD,2}}}} \right|}
		\\
		&=\frac{2}{F_{I}^{\left( 1 \right)}\left[ x_{LO} \right] N_{SD,2}N_{SD,1}}
		\\
		&\ \ \ \times \left| \sum_{i=1}^{N_{SD,1}}{\sum_{k=1}^{N_{SD,2}}{z_{k+\left( i-1 \right) \times N_{SD,2}}e^{-j\frac{2\pi \left( k-1 \right)}{N_{SD,2}}}}} \right|
		\\
		&=\frac{2}{N_{SD,2}N_{SD,1}F_{I}^{\left( 1 \right)}\left[ x_{LO} \right]} 
		\\
		&\ \ \ \times \left| \sum_{k=1}^{N_{SD,2}\times N_{SD,1}}{z_ke^{-j\frac{2\pi \left( k-1 \right)}{N_{SD,2}N_{SD,1}}N_{SD,1}}} \right|
		\\
		&=\frac{2}{NF_{I}^{\left( 1 \right)}\left[ x_{LO} \right]}\left| \sum_{k=1}^N{z_ke^{-j\frac{2\pi \left( k-1 \right)}{N}N_{SD,1}}} \right|.
	\end{aligned}
\end{equation}

\section{CRLB and Iteration Calculation of Splitting Detection}\label{A3}

\subsection{Iteration Calculation of Univariate Peak Shift Estimation}\label{A31}

Consider the optimum of $f_R$ must be a stationary point of Eq. (\ref{EqUE}), such that at the optimum $f_R$ we have $\frac{\partial}{\partial f_R}\ln p\left( \boldsymbol{z}_R;f_R \right) =0$, which is given by
\begin{equation}\label{EqUe_A1}
	\begin{aligned}
\sum_{i=1}^{N_{SF,1}}{\left( z_i-F_{SR}\left[ f_i-f_R \right] \right) \left( \left. \frac{\partial F_{SR}\left[ f \right]}{\partial f} \right|_{f=f_i-f_R} \right)}=0,
\end{aligned}
\end{equation}
where $\frac{\partial}{\partial f_R}F_{SR}\left[ f_i-f_R \right] =-\left. \frac{\partial F_{SR}\left[ f \right]}{\partial f} \right|_{f=f_i-f_R}$.

Given current estimate of shift $f_{R}^{\left[a \right] }$, neglecting higher order terms, we have
\begin{equation}\label{EqUe_A2}
	\begin{aligned}
&F_{SR}\left[ f_i-f_R \right] 
\\
&\ \ \ \ \ \ \ \ \ \approx F_{SR}\left[ f_i-f_{R}^{\left( a \right)} \right] -\left. \frac{\partial F_{SR}\left[ f \right]}{\partial f} \right|_{f=f_i-f_{R}^{\left[ a \right]}}\Delta f_{R}^{\left[ a \right]},
\end{aligned}
\end{equation}
where $\Delta f_{R}^{\left[ a \right]}$ is the difference from the current estimate $f_{R}^{\left[a \right] }$.

Finally, by substituting the expression from Eq. (\ref{EqUe_A1}) for the $F_{SR}\left[ f_i-f_R \right] $ term in Eq. (\ref{EqUe_A2}), we derive the recursive relation in Eq. (\ref{EqUE}).

\subsection{CRLB of Univariate Peak Shift Estimation}\label{A32}

Since all samples $\boldsymbol{z}_R=\left[ z_1,z_2,\cdots ,z_{N_{SF,1}} \right] $ are assumed to be independent, the overall likelihood for the entire set of observations is given by
\begin{equation}
	\begin{aligned}
		&p\left( \boldsymbol{z}_R;f_R \right) =\prod_{i=1}^{N_{SF,1}}{\frac{1}{\sqrt{2\pi \sigma _{SF}^{2}}}}
		\\
		&\ \ \ \ \ \ \ \ \ \ \ \ \ \ \ \ \ \ \ \times  \exp\left( -\sum_{i=1}^{N_{SF,1}}{\frac{\left( z_i-F_{SR}\left[ f_i-f_R \right] \right) ^2}{2\sigma _{SF}^{2}}}\right) ,
	\end{aligned}
\end{equation}
where $\sigma_{SF}^2=\frac{\sigma _{0}^{2}}{N_{SF,2}}$.

The first- and second-order partial derivatives with respect to $x$ can be obtained $\frac{\partial ^2}{\partial x^2}\ln p\left( \boldsymbol{\tilde{x}};x \right)$. Then, by taking the expectation with $\mathbb{E}\left\{ z_i-F_{SR}\left[ f_i-f_R \right] \right\} =0$ and $\mathbb{E}\left\{ \left( z_i-F_{SR}\left[ f_i-f_R \right] \right) ^2 \right\} =\frac{\sigma _{0}^{2}}{N_{SF,2}}$, Fisher information of superheterodyne detection model can be written as

\begin{equation}
\begin{aligned}
\mathbb{E}\left\{ \frac{\partial ^2}{\partial f_{R}^{2}}p\left( \boldsymbol{z}_R;f_R \right) \right\} =-N_{SF,2}\sum_{i=1}^{N_{SF,1}}{\frac{\left( F_{SR}^{\left( 1 \right)}\left[ f_i-f_R \right] \right) ^2}{\sigma _{0}^{2}}}.
\end{aligned}
\end{equation}

The CRLB of single right peak shift $f_{R}$ can be expressed as
\begin{equation}
\begin{aligned}
\text{CRLB}_{UR}=-\frac{1}{\mathbb{E}\left\{ \frac{\partial ^2}{\partial f_{R}^{2}}p\left( \boldsymbol{z}_R;f_R \right) \right\}}.
\end{aligned}
\end{equation}

Because the two peaks are independent of each other in Eq. (\ref{UE}), we have
\begin{equation}
	\begin{aligned}
		\text{CRLB}_{UE}&=\kappa ^2\left( \text{CRLB}_{UR}+\text{CRLB}_{UL} \right) 
		\\
		&= \frac{\kappa ^2\sigma _{0}^{2}}{N_{SF,2}\sum_{i=1}^{N_{SF,1}}{\left( F_{SR}^{\left( 1 \right)}\left[ f_i-f_R \right] \right) ^2}} 
		\\
		&\ \ \ + \frac{\kappa ^2\sigma _{0}^{2}}{N_{SF,2}\sum_{i=1}^{N_{SF,1}}{\left( F_{SL}^{\left( 1 \right)}\left[ f_i-f_L \right] \right) ^2}} ,
	\end{aligned}
\end{equation}
where $\text{CRLB}_{UR}$ and $\text{CRLB}_{UL}$ are the CRLBs for the two peak shifts.

\subsection{CRLB of Multivariate Peak Shift Estimation}\label{A33}

The Fisher information matrix $\text{J}\left( \boldsymbol{\theta } \right)$ for vector parameter estimation $\boldsymbol{\theta }=\left[ f_R,\boldsymbol{v} \right] $ is a symmetric $\left( M+1\right)  \times \left( M+1\right) $ matrix with elements defined by:

\begin{equation}
\begin{aligned}
\left[ \text{J}\left( \boldsymbol{\theta } \right) \right] _{i,j}=-\mathbb{E}\left\{ \frac{\partial ^2\ln p\left( \boldsymbol{z}_R;\boldsymbol{\theta } \right)}{\partial \theta _i\partial \theta _j} \right\} .
\end{aligned}
\end{equation}
The CRLB of single right peak shift $f_{R}$ can be expressed as
\begin{equation}
\begin{aligned}
\text{CRLB}_{MR}=\left[\text{J}^{-1}\left( \boldsymbol{\theta } \right) \right]_{1,1},
\end{aligned}
\end{equation}
where $f_R$ represents the first element of the vector parameter $\left[ \boldsymbol{\theta}\right] _1=f_R$. Consequently, the CRLB of $f_{R}$ corresponds to the first diagonal element of the inverse Fisher information matrix $\left[\text{J}^{-1}\left( \boldsymbol{\theta } \right) \right]_{1,1}$.

Because the two peaks are independent of each other in Eq. (\ref{ME}), we have
\begin{equation}
	\begin{aligned}
		\text{CRLB}_{ME}&=\kappa ^2\left( \text{CRLB}_{MR}+\text{CRLB}_{ML} \right) ,
	\end{aligned}
\end{equation}
where $\text{CRLB}_{MR}$ and $\text{CRLB}_{ML}$ are the CRLBs for the two peak shifts.

\ifCLASSOPTIONcaptionsoff
\newpage
\fi

\normalem
\bibliographystyle{IEEEtran}
\bibliography{myref}





%

%
%
%




\end{document}